# Superconducting and magnetic properties of a new EuAsFeO$_{0.85}$F$_{0.15}$ superconductor.

V.M.Dmitriev[1,2,*], I.E.Kostyleva[1,3], E.P.Khlybov[1,3], A.J.Zaleski[4], A.V.Terekhov[2], L.F.Rybaltchenko[2], E.V.Khristenko[2], L.A.Ishchenko[2].

1. International Laboratory for High Magnetic Fields and Low Temperatures, Gajowicka 95, 53-421 Wroclaw, Poland.

2. B.I.Verkin Institute for Low Temperature Physics and Engineering, NAS of Ukraine, 47 Lenin Ave., Kharkov, 61103, Ukraine.

3. L.F.Vereshchagin Institute for High-Pressure Physics, RAS, Troitsk, 142190, Russia.

4. W.Trzebiatowski Institute of Low Temperature and Structure Research, Polish Academy of Sciences, P.O.Box 1410, 50-950, Wroclaw, Poland.

## Abstract

Polycrystalline samples of a new superconducting EuAsFeO$_{0.85}$F$_{0.15}$ compound with critical temperature $T_c \approx 11$K were prepared by solid state synthesis. Its electric and magnetic properties have been investigated in magnetic fields from 0.1 to 140000 Oe. Critical magnetic fields $H_{c1}$, and $H_{c2}$ were measured and hence the magnetic penetration depths $\lambda$ and the coherence length $\xi$ have been estimated. The temperature dependence $H_{c2}$ (T) exhibits clear hyperbolic – type behavior starting with the lowest fields. The data derived were used to estimate probable high $T_c$ and $H_{c2}$ in compounds doped with rare-earths having small atomic radii.

**Introduction**

At the beginning of 2008, LaFeAsO$_{1-x}$F$_x$ was reported to be superconducting with Tc≈ 26K [1]. Soon afterwards, the critical temperature $T_c$ was raised to 40-43 K by substituting Ce [2] and Sm [3] for La and even to $T_c$ ≈ 52K with Nd and Pr [4,5] substitutions. Furthermore the superconducting transition temperature of the SmFeAsO$_{1-x}$F$_x$ samples, synthesized under high pressure (6 GPa), was $T_c$≈ 55 K [6]. Thus, there are good reasons for classifying the new group of compounds as high-$T_c$ superconductors. The case closely resembles the observations of $T_c$ increase in rare-earth REBaCuO systems. Also band-structure calculations and experiments actually point to a nontrivial mechanism of pairing in the new compounds (called pnictides). The presence of Fe and Pr antagonistic to traditional superconductivity in superconducting compounds strengthens the above conclusion.

The tetragonal crystalline structure of the new superconductors with the alternating La$_2$O$_{2-x}$F$_x$ and Fe$_2$As$_2$ layers [1] reminds HTSC structures. The FeAs layers are current-carrying and act much as the CuO$_2$ layers in cuprates, i.e. form the electron states near the Fermi surface. The LaOF layers are suppliers of charge carriers.
A number of four-component materials including Ce, Pr, Nb, Sm, Gd, Tb, Dy rare-earths have been synthesized lately by different groups. It was found that the highest $T_c$ is achievable in compounds with a proper content of fluorine (x ≈ 0.1-0.2). Moreover, the smaller the atomic radius of the rare-earth element, the higher $T_c$ is [3].

The goal of this study was to find out whether a rare-earth element with a large atomic radius would suppress the critical temperature $T_c$ of the REFeAsO$_{1-x}$F$_x$ compound. It is also interesting how this can affect the value and the temperature dependence of the upper critical magnetic field Hc$_2$. If Hc$_2$ is much below the literature data, the observation of the Hc$_2$ (T) behavior can be extended towards lower temperatures using a standard superconducting magnet.

As RE ion we have chosen Eu which atomic radius is 0.2023 nm.

The atomic radii of the rare-earths used hitherto are within 0.1755 - 0.1855 nm, the F-contents being 0.1-0.18, which comes as the optimal doping ensuring high $T_c$.

**Experimental details**

We prepared polycrystalline $EuAsFeO_{0.85}F_{0.15}$ material by ordinary solid state synthesis of EuAs, $EuF_3$, Fe and $Fe_2O_3$ compounds in an ampoule at $T=1150^0C$ for 24 hours. Additionally, post-grinding homogenization was performed at the same temperature for 30 hours.

The value and the temperature dependence of the electric resistance of the synthesized superconductors were investigated by the four-probe method in magnetic fields H up to 14 T on 5x1x1 mm samples cut out of tablets. A PPMS instrument was used to measure both the magnetic AC susceptibility in low fields up to 10Oe and the DC magnetization in magnetic fields up to 9T.

**Results and discussion**

The temperature dependence of electric resistivity near the superconducting transition in magnetic fields up to 14T is shown in Fig.1. The superconducting transition temperature found from the onset of the transition is $T_c^{onset} \approx 11.4$ K. Since the atomic radius of Eu is considerably larger than those of the rare-earths used by other researchers, our data retain the tendency evident in the literature. This tendency is illustrated in Fig.2 showing experimental $T_c$ as a function of the rare–earth atomic radius (solid symbols). A qualitative prediction is made concerning possible $T_c$- values with smaller radii rare–earths (light symbols).

The field dependence of resistivity at the temperatures specified at each curve is shown in Fig.3. The dashed horizontal lines mark the resistivity levels making 10%, 50% and 90% of the normal resistivity value $\rho_N$. Fig.4 illustrates three dependences of $H_{c2}$ (T), which correspond to the resistivity levels from Fig.3. The circles at the highest field curve are magnetic measurement data (see below). The inset shows the initial part of this curve. Unlike [8] all the three curves exhibit quite clear hyperbolic–type temperature dependences instead of parabolic ones typical of traditional, single-gap superconductors. We therefore believe that the comparison of these dependences on the basic of the criterion $H_{c2}(T=0)=-0.693T_c(\partial H_{c2}/\partial T)_{T=T_c}$ , developed for traditional superconductors with the parabolic dependence $H_{c2}(T)$, which is commonly accepted procedure [9], would be rather inadequate. On the other hand, near $T_c$ the $H_{c2}(T)$ – values are too high to extend measurement to the low temperature region, and the comparison of $H_{c2}(0)$ – values is often problematic for technical reasons. Therefore, in this study the results obtained on compounds with different rare- earths were compared using the $H_{c2}(T)$ – value measured at the same relative temperature $T/T_c = 0.95$ near $T_c$. The results are shown in Fig.5 as a function of the atomic radius of the rare – earth. It is seen that the $H_{c2}(T)$ – values, just as in the $T_c$ – case, increase with a decreasing atomic radius of rare-earths. We can thus expect that advent of advanced technologies, high magnetic fields at high $T_c$ are quite realistic in the new $REFeAsO_{1-x}F_x$ – type superconductor.

The temperature dependence of AC – magnetization $M'$(a), and $M''$ (b) in low fields up to 10 Oe and in fields up to 9T are shown in Figs.6 and 7, respectively. It is seen that even rather low magnetic fields have an appreciable effect on the superconducting transition temperature exhibiting the hyperbolic dependence of $H_{c2}$ (T), unusual for traditional superconductors (see the inset in Fig.4). This $H_{c2}(T)$ dependence can be caused by a multigap Fermi surface [12,13] or by some other factors resulting from complex charge and spin interactions. The $T_c^{onset}$ estimated from these data is ≈ 11 K. The $H_{c2}(T)$ dependence in Fig.4 obtained from the onset of the superconducting transition in these magnetic measurements (empty circles in Fig.4) are in good agreement with the results of resistance measurements. Fig.8 shows a curve at T=9 K based on the data of Figs.6 and 7 which describes the magnetization of our sample $EuAsFeO_{0.85}F_{0.15}$ in the fields up to 9 T. It is seen that the curve M(H) corresponds to the magnetization of a type II superconductor with a low critical magnetic field $H_{c1}$. The value of $H_{c1}$ at this temperature is $H_{c1}$≈10 Oe (see inset in Fig.8). We can thus estimate the magnetic penetration depth λ from the known expression $H_{c1} = (\Phi_0 /4\pi\lambda^2)\ln(\lambda/\xi)$, where $\Phi_0$ is the magnetic flux quantum, ξ is the coherence length. The value of ξ found from $H_{c2} = (\Phi_0 /2\pi\xi^2)$ at T=9K is equal to ξ(T=9K) ≈

60Å. As a result, $\lambda(T=9K) \approx 9000$Å. Then, the parameter $\kappa = \lambda/\xi \approx 150$. Thus, the new compounds are hard type II superconductors. In our opinion, standard extrapolation of $H_{c2}$ values to T=0 is unreasonable because the dependences $H_{c1}(T)$ and $H_{c2}(T)$ at $T \rightarrow 0$ are not known yet.

## Conclusions

In conclusion we note that one more pnictide – family superconducting compound $EuAsFeO_{0.85}F_{0.15}$ has been synthesized with $T_c \approx 11K$. Because of the large atomic radius of Eu, the compound $EuAsFeO_{0.85}F_{0.15}$ has lower $T_c$ and $H_{c2}$ in comparison with the known compounds. The results obtained permit us to predict superconducting compounds based on low atomic-radius rare-earths that can have both high $T_c$ and $H_{c2}$ simultaneously. Nevertheless, $H_{c2}$ in $EuAsFeO_{0.85}F_{0.15}$ is large enough to measure its complete temperature dependence with a standard 15T magnet. The dependence of $H_{c2}(T)$ is of the hyperbolic type even in very low fields (0.1-200Oe), which makes the WHH criterion [9] $H_{c2}(0)=-0.693T_c (\partial H_{c2}/\partial T)_{T-Tc}$ inadequate for estimating the upper critical field at T=0. The investigations of magnetization in weak fields enabled us to estimate the field $H_{c1}$, which is about 10 Oe at T=9K ($T/T_c \approx 0.8$). The coherence length $\xi(T=9K) \approx 60$Å, the magnetic penetration depth $\lambda(T=9K) \approx 9000$Å and the parameter $\kappa = \lambda/\xi \approx 150$ were also estimated for this temperature.

## Acknowledgment

Some of us thank the partially financial support of this study by the RFFI, grant № 08-08-00709-a.

**Figure Captions**

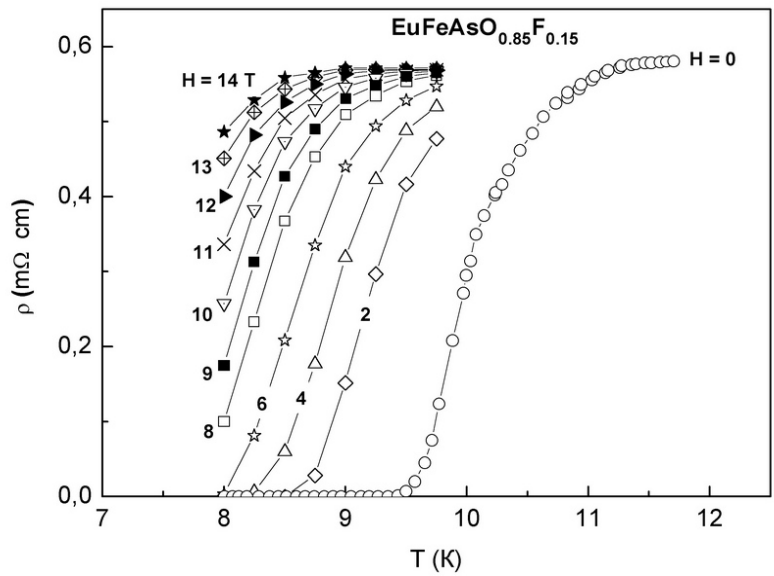

Fig.1. The temperature dependences of the resistivity of $EuAsFeO_{0.85}F_{0.15}$ near the superconducting transition in magnetic fields 0-14T (fields are specified at each curve).

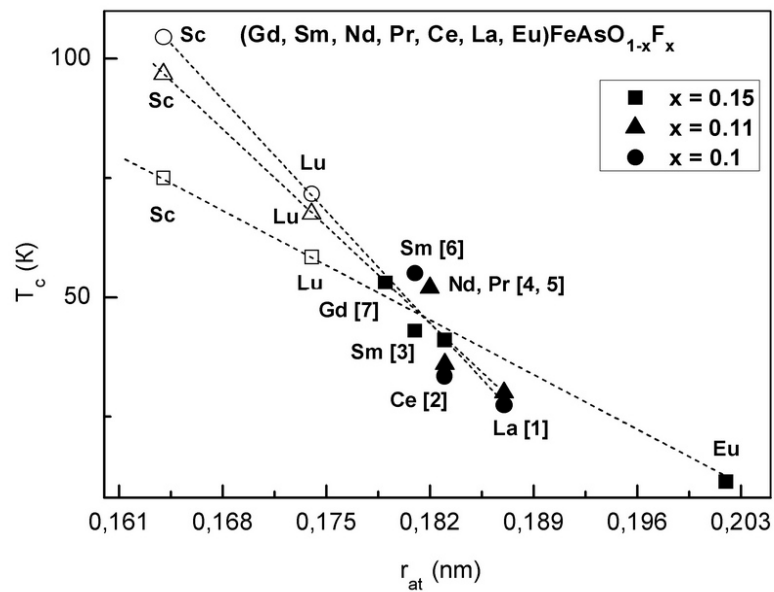

Fig.2. The dependence of the critical temperature $T_c$ upon the atomic radius $r_{at}$ of the rare – earth element in $REAsFeO_{1-x}F_x$ compounds (x=0.1, 0.11, 0.15). Solid symbols are for experimental results of the studies cited and this work. Open symbols are predictions for rare-earths with smaller atomic radii.

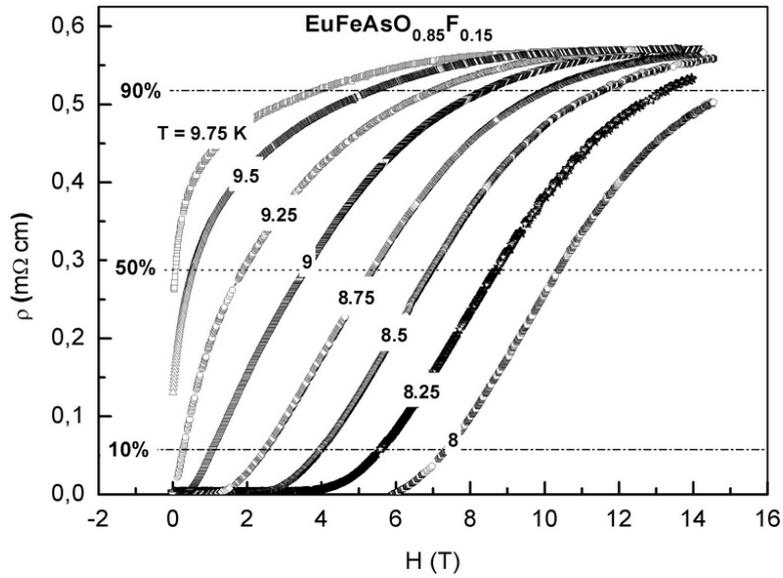

Fig.3. The magnetic field dependences of the resistivity ρ(H) in EuAsFeO$_{0.85}$F$_{0.15}$ at temperatures (shown at each curve) below the superconducting transition temperature. Dashed horizontal lines are ρ(H) – values at the levels making 10%, 50% and 90% of the normal resistivity ρ$_N$ at T ≥ T$_c$.

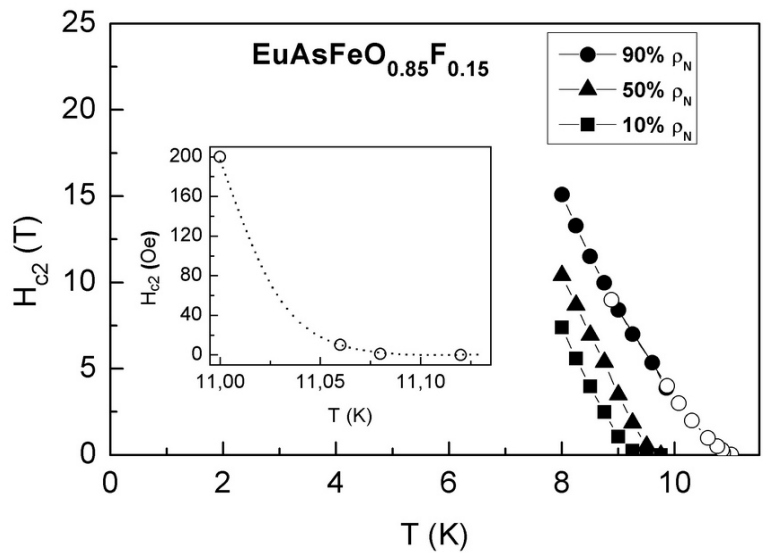

Fig.4. The temperature dependences of the second critical field H$_{c2}$ for EuAsFeO$_{0.85}$F$_{0.15}$ plotted by the data of Fig.3 for the sections of the dependences ρ(H) at the levels corresponding to 90%, 50% and 10% of ρ$_N$. Solid and open symbols indicate resistive and magnetic measurements, respectively. Inset: H$_{c2}$(T) in low magnetic fields (magnetic measurement).

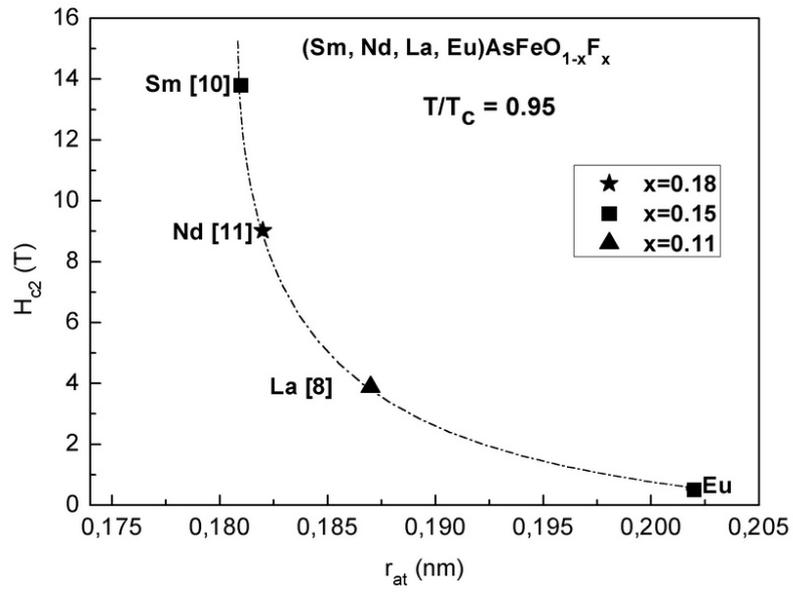

Fig.5. Experimental dependences of $H_{c2}$ on the rare–earths atomic radius $r_{at}$ in $REAsFeO_{1-x}F_x$ with x=0.11, 0.15, 0.18 at $T/T_c$ = 0.95. In brackets: literature sources.

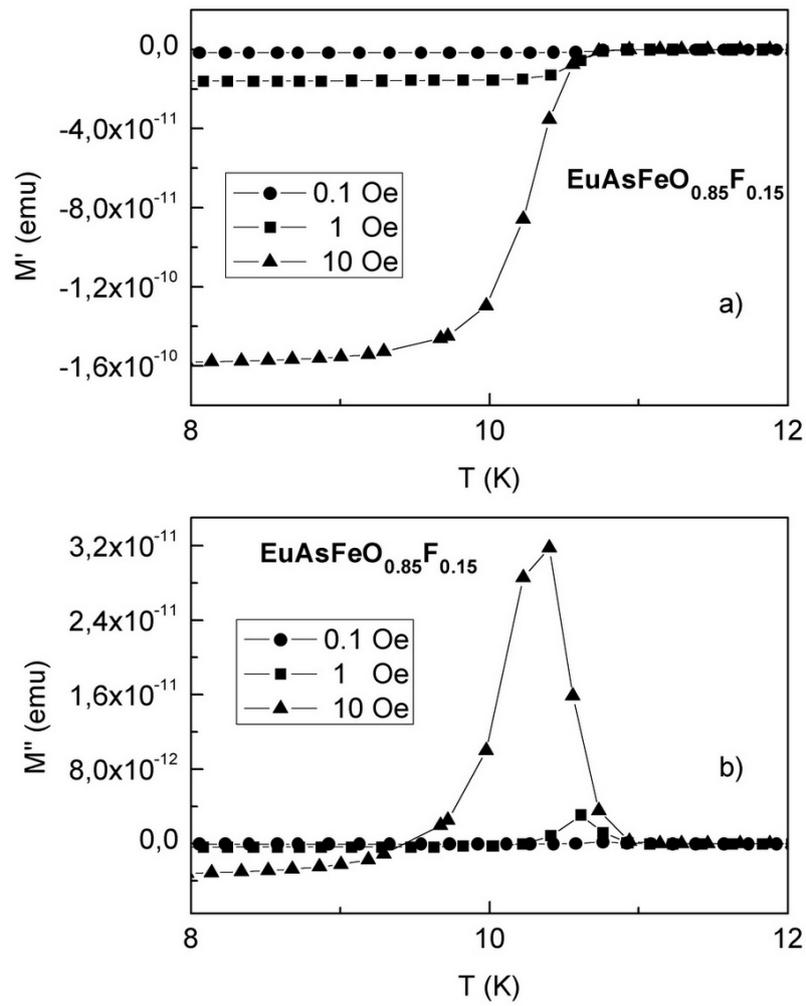

Fig.6. The temperature dependences of the AC – magnetization M′(a) and M″(b) in low magnetic fields up to 10 Oe.

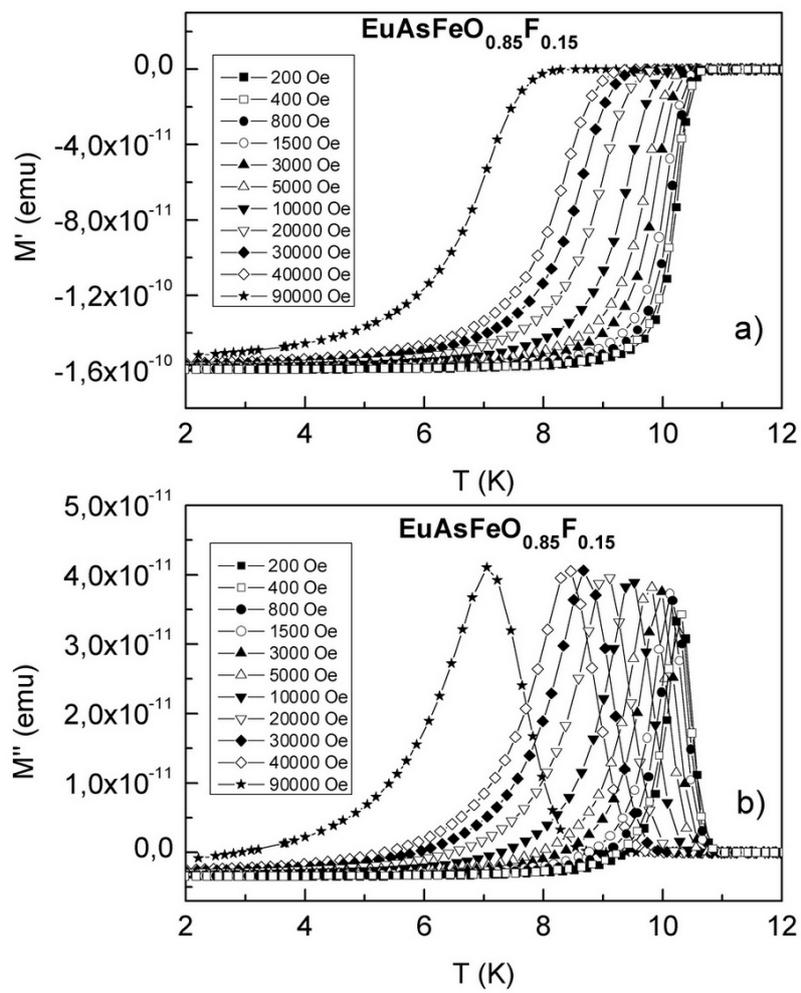

Fig.7. The temperature dependences of the AC – magnetization M′(a) and M″(b) in magnetic fields up to 9T.

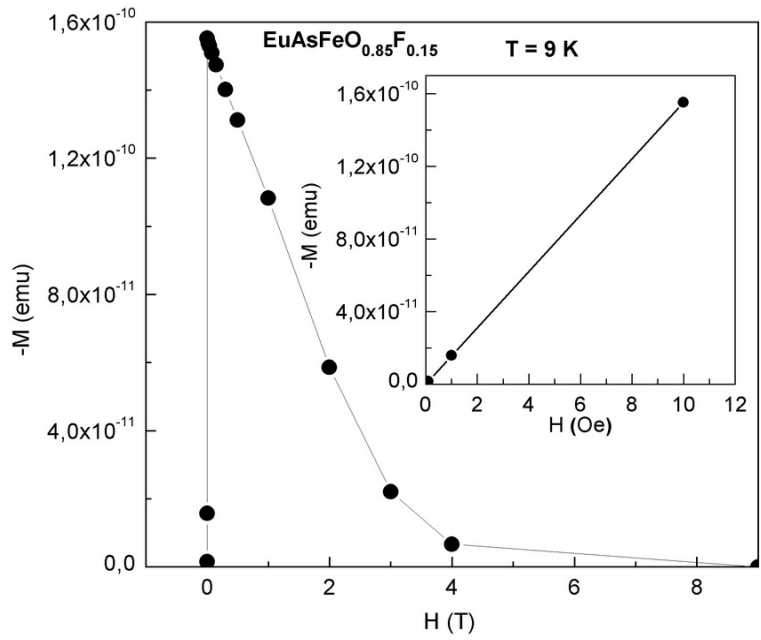

Fig.8. The magnetization curve of EuAsFeO$_{0.85}$F$_{0.15}$ in magnetic fields up to 9T at T=9K (from the data of Figs.6, 7). Inset: the initial part of the dependence M(H) in low magnetic fields.